\documentstyle[12pt]{article}

\renewcommand{\(}{$\displaystyle}
\renewcommand{\)}{$}

\newcommand{\ds}{\displaystyle}
\newcommand{\Tr}{\mbox{\rm Tr}}

\title{On validity of perturbative quantization of the breathing mode
in the Skyrme model}

\author{
A.~Kostyuk and A.~Kobushkin \\
Bogolyubov Institute for Theoretical Physics, \\
Ukrainian Academy of Sciences,\\
Kiev 143, Ukraine \\
\and N.~Chepilko
\\Institute of Physics,\\
Ukrainian Academy of Sciences, \\
Kiev 028, Ukraine \\
\and T.Okazaki \\
Physics Laboratory, Sapporo Campus,\\
Hokkaido University of Education,\\
Sapporo 002, Japan
}

\date{December 20, 1993}

\begin{document}
\maketitle

\begin{abstract}
We present a detailed discussion of the breathing mode quantization
in the Skyrme model and demonstrate that the chiral angle of the
hedgehog soliton is strongly affected by the breathing motion.
\end{abstract}

It was already demonstrated that the centrifugal effects seriously
change chiral angle (profile function) of a rotating skyrmion
\cite{Rajar}. At the same time the breathing mode in the Skyrme model
is usually studied under the assumption that its influence is small
enough for using perturbative quantization and calculations can be
done with the chiral angle of the static soliton \cite{Bied,Breath}.
In this Brief Report we investigate validity of this assumption and
show that the breathing motion affects strongly the soliton chiral
angle behavior.

We are starting from the standard Lagrangian of the model
\begin{equation}
{\cal L}(U) = - \frac{F_{\pi}^{2}}{16}
\Tr \left( L_{\mu}L^{\mu} \right)
+\frac{1}{32e_{S}^{2}} \Tr \left(
\left[ L_{\mu},L_{\nu} \right] \left[ L^{\mu},L^{\nu} \right]
\right)
+\frac{1}{16}F_{\pi}^{2}m_{\pi}^{2} \Tr(U+U^{+}-2),
\label{l-ian}
\end{equation}
where \( L_{\mu} = U^{+} \frac{\partial U} {\partial x^{\mu}} \);
$U=U(t,\vec{x})$ is the $SU(2)$ chiral field matrix;
$F_{\pi}$ and $m_{\pi}$ stand for the pion decay constant and pion
mass, respectively; $e_{S}$ is the dimensionless Skyrme constant.

The model has the well-known hedgehog static solution
with topological charge $B=1$ \cite{Adkins}
\begin{equation}
U_{0}=\exp \left[ i \vec{\tau} \hat{x} \theta_{0} (\tilde{r}) \right],
\label{h-hog}
\end{equation}
where $\vec{\tau}=(\tau_{1},\tau_{2},\tau_{3})$ are the
isotopic Pauli matrices, \( \hat x =\frac{\vec{x}}{|\vec{x}|} \),
 \( \tilde{r}=e_{S}F_{\pi}|\vec{x}| \). The chiral angle
$\theta_{0}(\tilde{r})$ is obtained from the variational principle
\begin{equation}
\frac{\delta\int{\cal L}(U_{0})d^{3}x}
{\delta\theta_{0}(\tilde{r})}=0
\label{varcl}
\end{equation}
supplemented by the boundary conditions
\begin{equation}
\left. \theta_{0}(\tilde{r}) \right|_{\tilde{r}=0} =\pi,
\label{bound0s}
\end{equation}
\begin{equation}
\left. \theta_{0}(\tilde{r}) \right|_{\tilde{r} \rightarrow
\infty}=0.
\label{boundinfs}
\end{equation}
Eq.(\ref{varcl}) is equivalent to the following differential equation
\begin{eqnarray}
 & & (\tilde{r}^2+8\sin^{2}\theta_{0})
\frac{d^2\theta_{0}}{d\tilde{r}^{2}}
+2\tilde{r}\frac{d\theta_{0}}{d\tilde{r}}  \nonumber \\
& &-\left( 1 + 4 \frac{\sin^{2}\theta_{0}}{\tilde{r}^{2}} -
4\left(\frac{d\theta_{0}}{d\tilde{r}}\right)^2 \right)
\sin 2\theta_{0} - \frac{m_{\pi}^{2}}{e_{S}^{2}F_{\pi}^{2}}
\tilde{r}^2 \sin \theta_{0} = 0.
\label{static}
\end{eqnarray}

Now let us consider the breathing and rotating hedgehog ansatz
\cite{Bied}
\begin{equation}
U=A(t) \exp \left( i \vec{\tau} \hat{x} \theta ( e^{\lambda(t)}
\tilde{r}) \right) A^{+}(t),
\label{h-hogm}
\end{equation}
where $A \in SU(2)$, $A=a_{0}(t)+i\vec{\tau}\vec{a}(t)$,
$a_{0}^{2}+\vec{a}^{2}=1$. The quantities $a_{p}=(a_{0},\vec{a})$,
$p=0,1,2,3$ and the homogeneous scale transformation parameter
$\lambda(t)$ are the quantum collective coordinates which describe
rotation and vibration of the soliton, respectively.
In terms of these collective coordinates the Lagrangian of the
dynamical system is written as \cite{Bied}:
\begin{equation}
L = \int d^{3}x {\cal L}(U) =
\frac{1}{2}A(\lambda) \dot{\lambda}^{2} - B(\lambda)
+\frac{1}{2}C(\lambda) \Tr (\dot{A}\dot{A}^{+}),
\label{l-func}
\end{equation}
where
\begin{eqnarray}
A(\lambda) &=& e^{-3\lambda}Q_{2}+e^{-\lambda}Q_{4},
\label{a} \\
B(\lambda) &=& e^{-\lambda}V_{2}+e^{\lambda}V_{4}
+e^{-3\lambda}V_{\pi},
\label{b} \\
C(\lambda) &=& e^{-3\lambda}I_{2}+e^{-\lambda}I_{4}
\label{c}
\end{eqnarray}
and
\begin{eqnarray}
Q_{2} &=& \frac{\pi}{e_{S}^{3}F_{\pi}} \int_{0}^{\infty} d\tilde{r}
\tilde{r}^{4} \left(\frac{d\theta(\tilde{r})}{d\tilde{r}}\right)^{2},
\label{q2} \\
Q_{4} &=& \frac{8\pi}{e_{S}^{3}F_{\pi}} \int_{0}^{\infty} d\tilde{r}
\tilde{r}^{2}\left(\frac{d\theta(\tilde{r})}{d\tilde{r}}\right)^{2}
\sin^{2}\theta(\tilde{r}),
\label{q4} \\
V_{2} &=& \frac{\pi F_{\pi}}{2e_{S}}\int_{0}^{\infty} d\tilde{r}
\tilde{r}^{2} \left( \left(\frac{d\theta(\tilde{r})}{d\tilde{r}}
\right)^{2} +\frac{2sin^{2}\theta(\tilde{r})}{\tilde{r}^{2}} \right),
\label{v2} \\
V_{4} &=& \frac{2\pi F_{\pi}}{e_{S}} \int_{0}^{\infty} d\tilde{r}
\sin^{2}\theta(\tilde{r}) \left(
2\left(\frac{d\theta(\tilde{r})}{d\tilde{r}}\right)^{2}
+\frac{\sin^{2}\theta(\tilde{r})}{\tilde{r}^{2}} \right),
\label{v4} \\
V_{\pi} &=& \frac{2\pi m_{\pi}^{2}}{F_{\pi}e_{S}^{3}}
\int_{0}^{\infty} d\tilde{r} \tilde{r}^{2}
\sin^{2}\frac{\theta}{2},
\label{vpi} \\
I_{2} &=& \frac{4\pi}{3F_{\pi}e_{S}^{3}} \int_{0}^{\infty} d\tilde{r}
\tilde{r}^{2} \sin^{2}\theta(\tilde{r}),
\label{i2} \\
I_{4} &=& \frac{16\pi}{3F_{\pi}e_{S}^{3}} \int_{0}^{\infty} d\tilde{r}
\tilde{r}^{2} \sin^{2} \theta(\tilde{r})
\left( \left(\frac{d\theta(\tilde{r})}{d\tilde{r}}\right)^{2}+
\frac{\sin^{2}\theta(\tilde{r})}{\tilde{r}^{2}} \right).
\label{i4}
\end{eqnarray}
(Integrals with the subscripts $2$, $4$ and $\pi$ are the
contributions of the kinetic term, the Skyrme term and the symmetry
breaking term, respectively.)

In the perturbative quantization approach  \cite{Bied,Breath}
the integrals (\ref{q2})-(\ref{i4}) are estimated by means of
substitution of the static solution $\theta_{0}(\tilde{r})$
instead of $\theta(\tilde{r})$.

To take into account the influence of the breathing mode
nonperturbatively, one has to determine the chiral angle
$\theta(\tilde{r})$ from the variational principle averaged
over the quantum state $|nj\rangle $ \cite{Scale}:
\begin{equation}
\left\langle nj \left| \frac{\delta
L}{\delta \theta(\tilde{r})} \right| nj \right\rangle = 0 ,
\label{varqu}
\end{equation}
where  $n$ and $j$ denote quantum numbers of the rotational and
breathing excitations, respectively.

Using the  following notations
\begin{eqnarray}
\alpha_{2}^{2} &=& \frac{1}{e_{S}^{2}F_{\pi}^{2}} \frac{\langle
nj|\dot{\lambda}^{2}e^{-3\lambda }|nj\rangle }
{\langle nj|e^{-\lambda}|nj\rangle },
\label{alf2}\\
\alpha_{4}^{2} &=& \frac{4}{e_{S}^{2}F_{\pi}^{2}}
\frac{\langle nj|\dot{\lambda}^{2}e^{-\lambda }|nj\rangle }
{\langle nj|e^{-\lambda }|nj\rangle },
\label{alf4}\\
\beta_{4}^{2} &=& \frac{\langle nj|e^{\lambda }|nj\rangle }
{\langle nj|e^{-\lambda }|nj\rangle },
\label{bet4}\\
\beta_{\pi }^{2} &=& \frac{m_{\pi }^{2}}{e_{S}^{2}F_{\pi}^{2}}
\frac{\langle nj|e^{-3\lambda }|nj\rangle}
{\langle nj|e^{-\lambda }|nj\rangle },
\label{betpi}\\
\gamma_{2}^{2} &=& \frac{2}{3e_{S}^{2}F_{\pi}^{2}}
\frac{\langle nj|e^{-3\lambda }\Tr(\dot{A}\dot{A}^{+})|nj\rangle }
{\langle nj|e^{-\lambda }|nj\rangle },
\label{gam2}\\
\gamma_{4}^{2} &=& \frac{8}{3e_{S}^{2}F_{\pi}^{2}}
\frac{\langle nj|e^{-\lambda}\Tr(\dot{A}\dot{A}^{+})|nj\rangle }
{\langle nj|e^{-\lambda}|nj\rangle },
\label{gam4}
\end{eqnarray}
one can rewrite (\ref{varqu}) in the form\footnote{
Eq.(\ref{breath}) can also be obtained from the minimization condition
on the total energy, while the solution of Eq.(\ref{static}) minimizes
only the skyrmion static energy.}
\begin{eqnarray}
& &
\left(\tilde{r}^{2}(1-\alpha_{2}^{2}\tilde{r}^{2})+
\left(8\beta_{4}^{2}-2(\alpha_{4}^{2}+
\gamma_{4}^{2})\tilde{r}^{2}\right)\sin^{2}\theta \right)
\frac{d^{2}\theta }{d\tilde{r}^{2}} \nonumber \\
& & +\left(2\tilde{r}(1-2\alpha_{2}^{2}\tilde{r}^{2})-
4(\alpha_{4}^{2}+\gamma_{4}^{2})\tilde{r}\sin^{2}\theta \right)
\frac{d\theta }{d\tilde{r}} \nonumber \\
& &+\left(4\beta_{4}^{2}-(\alpha_{4}^{2}+\gamma_{4}^{2})\tilde{r}^{2}
\right)\left(\frac{d\theta }{d\tilde{r}}\right)^{2} \sin 2\theta
\label{breath} \\
& &-\left(1-\gamma_{2}^{2}\tilde{r}^{2}+
\left(\frac{4\beta_{4}^{2}}{\tilde{r}^{2}}
-2\gamma_{4}^{2}\right)\sin^{2}\theta \right) \sin 2\theta
-\beta_{\pi}^{2}\tilde{r}^{2} \sin \theta = 0.  \nonumber
\end{eqnarray}
It should be supplemented by the boundary conditions
\begin{equation}
\left. \theta(\tilde{r}) \right|_{\tilde{r}=0} =\pi,
\label{bound0}
\end{equation}
\begin{equation}
\left. \theta(\tilde{r}) \right|_{\tilde{r} \rightarrow
\infty}=0.
\label{boundinf}
\end{equation}

The main difference between the static case and the quantum
breathing one consists in the factors of the second order
derivative terms in Eqs.(\ref{static}) and (\ref{breath}):
contrary to (\ref{static}), in Eq.(\ref{breath}) this factor
contains the term proportional to $\tilde{r}^{4}$. The latter arises
from the breathing motion kinetic energy. This difference changes the
asymptotic behavior of the chiral angle drastically.
Let us compare the asymptotic solutions of the two equations
at large $\tilde{r}$. First,  Eq.(\ref{static}) is asymptotically
reduced to
\begin{equation}
\tilde{r}^{2}\frac{d^{2}\theta_{0}^{\infty}}{d\tilde{r}^{2}}
+2\tilde{r}\frac{d\theta_{0}^{\infty} }{d\tilde{r}}
-(2+\mu_{\pi }^{2}\tilde{r}^{2})\theta_{0}^{\infty} = 0,
\label{asstat}
\end{equation}
where $\displaystyle \mu _{\pi }^{2}=
\frac{m_{\pi }^{2}}{e_{S}^{2}F_{\pi }^{2}}$ and
$\theta_{0}^{\infty}(\tilde{r}) = {\lim_{r\to\infty}\theta_{0}(\tilde{r})}
\ll 1.$
The general solution of Eq.(\ref{asstat}) is
\begin{equation}
\theta _{0}^{\infty}=C_{1}\left(\frac{\mu _{\pi }}{\tilde{r}}-
\frac{1}{\tilde{r}^{2}}\right) e^{\mu _{\pi }\tilde{r}}
+C_{2}\left(\frac{\mu _{\pi }}{\tilde{r}}+
\frac{1}{\tilde{r}^{2}}\right) e^{-\mu _{\pi }\tilde{r}}.
\label{genstat}
\end{equation}
In order to satisfy the boundary condition (\ref{boundinfs}) the coefficient
$C_{1}$ in (\ref{genstat}) must be zero which can be realized by
appropriate choice of $\theta'(\tilde{r})|_{\tilde{r}=0}$.

Next, the asymptotic form of Eq.(\ref{breath}) is
\begin{equation}
\tilde{r}^{2}\frac{d^{2}\theta^{\infty} }{d\tilde{r}^{2}}
+4\tilde{r}\frac{d\theta^{\infty} }{d\tilde{r}}-\nu \theta^{\infty} =0
\label{asbreath}
\end{equation}
with $\displaystyle \nu =
\frac{2\gamma _{2}^{2}-\beta _{\pi }^{2}}{\alpha _{2}^{2}}$,
and the general solution is expressed as \cite{Ours}
\begin{equation}
\theta^{\infty}(\tilde{r}) = \left\{
\begin{array}{lcr}
\ds C_{1} \tilde{r}^{\kappa_{1}} +  C_{2} \tilde{r}^{\kappa_{2}},
&\ds {\rm if} &\ds \nu >-\frac{9}{4} \\[1.5ex]
\ds C_{1} \tilde{r}^{-3/2} \ln \tilde{r} +  C_{2} \tilde{r}^{-3/2},
&\ds {\rm if} &\ds \nu=-\frac{9}{4} \\[1.5ex]
\ds C_{1} \tilde{r}^{-3/2} \sin \left(\kappa_{0}\ln \tilde{r} \right)
+ C_{2} \tilde{r}^{-3/2} \cos \left(\kappa_{0}\ln \tilde{r} \right),
&\ds {\rm if} &\ds \nu<-\frac{9}{4}
\end{array} \right.
\label{asymptsol}
\end{equation}
where
\begin{equation}
\kappa_{1,2} = -\frac{3}{2} \pm
\sqrt{\left(\frac{3}{2}\right)^{2}+\nu} ,
\mbox{\hspace{3em}}
 \kappa_{0} = \sqrt{|\nu|-\left(\frac{3}{2}\right)^{2}} .
\label{kappa120}
\end{equation}

First of all, we have to note that (contrary to the static case)
the coefficient $C_{1}$ in Eq.(\ref{asymptsol}) cannot be
chosen equal zero by appropriate choice of
$\theta'(\tilde{r})|_{\tilde{r}=0}$ due to the following
reason.  The factor at second order derivative of Eq.(\ref{breath})
\begin{equation}
g(\tilde{r})=
\tilde{r}^{2}(1-\alpha_{2}^{2}\tilde{r}^{2})+
\left(8\beta_{4}^{2}-2(\alpha_{4}^{2}+
\gamma_{4}^{2})\tilde{r}^{2}\right)\sin^{2}\theta
\label{factor}
\end{equation}
is positive for small $\tilde{r}$ and becomes negative for
large $\tilde{r}$, so there exists at least one point
$\tilde{r}_{1}$, where $g(\tilde{r}_{1})=0$. This point is a singular
point of Eq.(\ref{breath}) and in order to get regular solution
the following condition must be fulfilled
\begin{eqnarray}
& & \left. \left(2\tilde{r}(1-2\alpha_{2}^{2}\tilde{r}^{2})-
4(\alpha_{4}^{2}+\gamma_{4}^{2})\tilde{r}\sin^{2}\theta \right)
\frac{d\theta }{d\tilde{r}}\right|_{\tilde{r}=\tilde{r}_{1}}
\nonumber \\
& & +\left.\left(4\beta_{4}^{2}+(\alpha_{4}^{2}
+\gamma_{4}^{2})\tilde{r}^{2}
\right) \left(\frac{d\theta }{d\tilde{r}}\right)^{2} \sin 2\theta
\right|_{\tilde{r}=\tilde{r}_{1}}
\label{regular} \\
& &-\left.\left(1-\gamma_{2}^{2}\tilde{r}^{2}+
\left(\frac{4\beta_{4}^{2}}{\tilde{r}^{2}}
-\gamma_{4}^{2}\right)\sin^{2}\theta \right) \sin 2\theta
\right|_{\tilde{r}=\tilde{r}_{1}}
-\left.\beta_{\pi}^{2}\tilde{r}^{2} \sin \theta
\right|_{\tilde{r}=\tilde{r}_{1}} = 0 . \nonumber
\end{eqnarray}
A choice of $\theta'(\tilde{r})|_{\tilde{r}=0}$ is
restricted by the fulfillment of the condition (\ref{regular}).
Assuming, in addition, the fulfillment of the condition $C_{1}=0$,
one obtains an overdefined boundary-value problem (the differential
equation of the second order (\ref{breath}) supplemented by the three
conditions (\ref{bound0}), (\ref{regular}) and $C_{1}=0$). Such a problem
has a solution only for a special choice of some spectral parameter. But
in the case under consideration the problem has no such parameter:
all coefficients of Eq.(\ref{breath}) are fixed by
Eqs.(\ref{alf2}-\ref{gam4}). Thus {\em regular} solution
satisfying (\ref{bound0}) does not satisfy $C_{1}=0$.

It is easily seen that for the case $\kappa _{1} \geq 0$ (which
corresponds to $\nu \geq 0$ ) the condition (\ref{boundinf}) is not
fulfilled unless $C_{1}=0$. At $\nu < 0$ the condition
(\ref{boundinf}) can be fulfilled for any $C_{1}$ and $C_{2}$, but
there remain some problems about the functionals $Q_{2}$, $V_{\pi}$
and $I_{2}$. If $\displaystyle \nu \leq -\frac{9}{4}$ they are
divergent for arbitrary nontrivial values of $C_{1}$ and $C_{2}$.
When $\displaystyle -\frac{9}{4}<\nu <0$, $C_{1}$ has to be zero in
order to get finite values of these functionals.
So there is no solution for the case of $\nu \geq 0$ (such situation
arises, for instance, in the chiral limit ($m_{\pi}=0$) ), when
$\nu <0$ the existence of solution is not ruled out, but the
integrals $Q_{2}$, $V_{\pi}$ and $I_{2}$ should be divergent.

It was shown that the nonperturbative quantization of the breathing
and rotating modes gives stable soliton solutions in the nonlinear
$\sigma$-model {\em without} the Skyrme term \cite{Ours}.
Their energy and mean square radii are finite in spite of
divergency of the functionals $Q_{2}$, $V_{\pi }$ and $I_{2}$.
One can expect that in the present model solutions with finite
energy and size could also exist.
But, in any case, the slow asymptotic decreasing (at $ \displaystyle
-\frac{9}{4}\leq \nu <0$) and oscillating behavior (at
$\displaystyle \nu < -\frac{9}{4}$) of the chiral angle
at large $\tilde{r}$ contradicts the Yukawa law. This problem
inherent in the rotating soliton ($\gamma_{2},
\gamma_{4} \not= 0$) as well as in nonrotating one ($\gamma_{2},
\gamma_{4} = 0$) what means that the source of these difficulties is
in the assumption about homogeneous global breathing which does not
describe the soliton external part correctly.

Our main result implies that the nonperturbative quantization of
the homogeneous global breathing of the skyrmion gives the solutions
different drastically from the static ones.
In this case the using of the perturbative approach is problematic.
One would also face the similar problems considering  the
homogeneous global breathing quantization of multiskyrmions, solitons in
modified Skyrme model and the global nonspherically-symmetrical time
dependent deformation of skyrmions \cite{Modif}.

\end{document}